\documentclass[a4paper,11pt]{article}
\pdfoutput=1 
\usepackage{jinstpub} 
\usepackage{subfigure}
\usepackage{mathtools}
\usepackage{paralist}       
\usepackage{textgreek}
\usepackage{graphicx}

\usepackage{siunitx}

\usepackage{textcomp}

\title{\boldmath A sparkless resistive glass correction electrode for the spherical proportional counter }

\author[a,1]{I. Katsioulas\note{Corresponding author.},}
\author[a]{I. Giomataris,}
\author[a,b]{P. Knights,}
\author[a]{M. Gros,}
\author[a]{X.F. Navick,}
\author[b]{K. Nikolopoulos,}
\author[c]{I. Savvidis}

\affiliation[a]{IRFU, CEA, Universite Paris-Saclay, F-91191 Gif-sur-Yvette, France}
\affiliation[b]{School of Physics and Astronomy, University of Birmingham, B15 2TT, United Kingdom}
\affiliation[c]{Aristotle University of Thessaloniki, Greece }

\emailAdd{ioannis.katsioulas@cea.fr}

\abstract{
A new anode support structure for the spherical proportional counter is presented that incorporates a resistive correction electrode made of glass. This electrode improves the electric field homogeneity versus angle, while suppressing the probability and intensity of sparks compared to non-resistive alternatives. The configuration of the correction electrode was optimised with simulations. 
Such support structures have been constructed and measurements have demonstrated homogeneous response of the detector and operational stability.
A measurement of the resistivity of the glass used is also presented.  
}

\keywords{Electronic detector readout concepts, Detector modelling and simulations II, Gaseous detectors, Materials for gaseous detectors, Dark Matter detectors, Very low-energy charged particle detectors}

\begin{document}
\maketitle
\flushbottom

\section{Introduction}
\label{sec:intro}

The Spherical Proportional Counter~\cite{Giomataris:2005bb,Giomataris2008-mx} is a gaseous detector comprising a grounded spherical metallic shell acting as the cathode, and a small concentric sphere, the anode. The anode is supported by a grounded metallic rod. A wire through the centre of the rod supplies the anode high-voltage and the detector read-out. A review of recent developments in and applications of the spherical proportional counter is given in Ref.~\cite{Arnaud:2018bpc}.

\begin{figure}[htbp]
\centering 
\includegraphics[width=.4\textwidth]{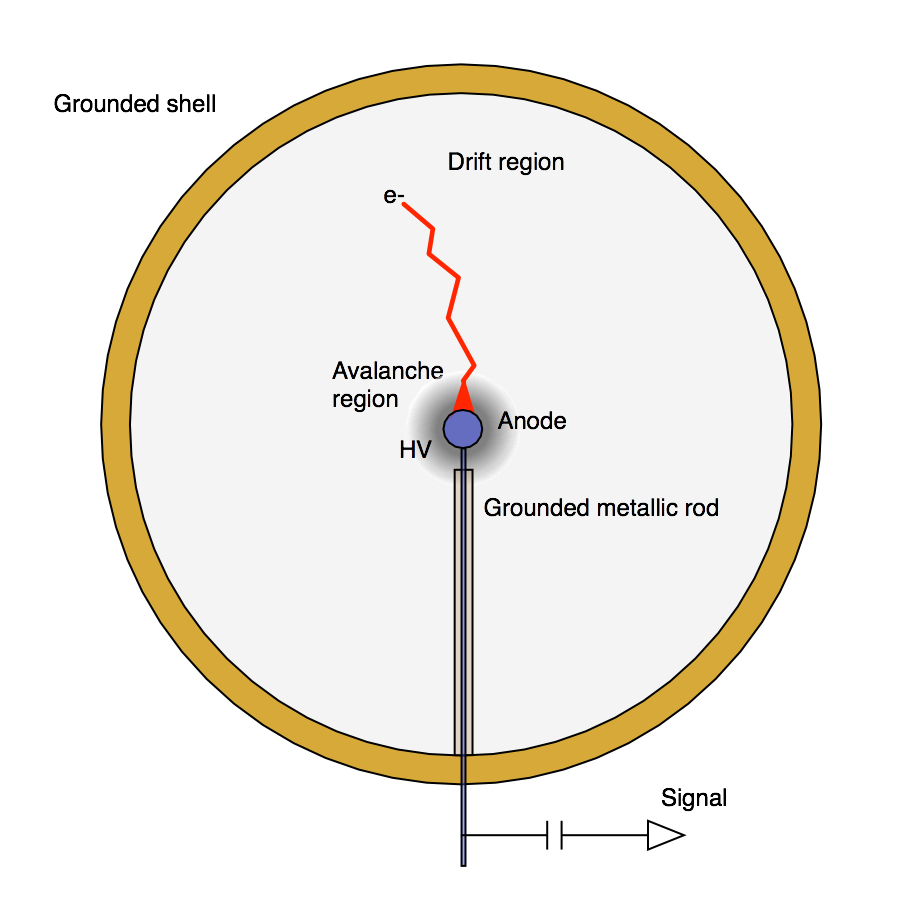}
\caption{\label{fig:1} Schematic of a spherical proportional counter displaying the principle of operation.}
\vspace{-0.5cm}
\end{figure}

For the ideal case of a floating anode of radius $r_1$ at a voltage $V_{0}$ and located at the centre of a spherical cathode shell of radius $r_2$, the magnitude of the radial electric field in the volume is
\begin{align}
E(r) & = \frac{V_{0}}{r^{2}}\frac{r_{1}\cdot r_{2}}{r_{2}-r_{1}}
\nonumber
\\
\xRightarrow{r_1\ll r_2}E(r) & \approx \frac{V_{0}}{r^{2}}r_{1}\; .
\label{eq:efield:2}
\end{align}
Equation~\ref{eq:efield:2} holds to a good approximation given that typically anode radii vary between $6~{\rm mm}$ and $0.5~{\rm mm}$, with increasing gas pressure,
while cathode radii are of $O(10~{\rm cm})$ or more.
The $r^{-2}$ dependence of the electric field naturally divides the detector into two regions: the drift and the multiplication volume.
This is shown in Fig.~\ref{fig:1}, where electrons liberated in the drift volume will drift under the influence of the weak electric field towards the anode. Once a few millimetres from the anode, under the influence of the strong electric field, an avalanche will form. This dependence also leads to the drift time dispersion, $\sigma_t(r)$, of the initial electrons to be radius-dependent: $\sigma_t(r)\sim(r/r_{2})^{3}$, a property exploited for detector fiducialisation. Furthermore, the difference in electron drift time between events with a spatially extended energy deposit versus those with a point-like energy deposit provides a powerful handle for background rejection~\cite{Bougamont2017-km, Savvidis2018-xo, Arnaud2018-nr}. 
Another feature of the detector is the very low capacitance, which is also practically independent of the detector volume since for $r_1\ll r_2$ it is given as $C\approx4\pi\epsilon\epsilon_0 r_1$.
This feature together with the large achievable gain~\cite{Iguaz_Gutierrez2015-ix} allows the development of large spherical proportional counters with an energy threshold as low as a single electron~\cite{Bougamont2010-xp}. 

These features suggest that the spherical proportional counter could be a good candidate for light dark matter searches, in the mass range from $0.1~{\rm GeV}$ to $10\;{\rm GeV}$~\cite{Arnaud2018-nr}, a possibility explored by the NEWS-G collaboration~\cite{news-g_web}. The detection relies on the measurement of low energy recoils, keV or lower, induced in the detector volume by elastic scattering of dark matter particles on the target nuclei. In this context, using detectors with sub-keV energy threshold gives a significant advantage, as discussed extensively in the literature~\cite{Agnese2014-ln,DAMIC_Collaboration2016-di,The_CRESST_Collaboration2015-pz}. The possibility to use various light targets, such as hydrogen, helium and neon, by changing the gas composition is also important, to maximize the recoil energy.

The NEWS-G Collaboration operates a $60~{\rm cm}$ diameter detector in the Laboratoire Souterrain de Modane~\cite{lsm}, which already yielded constraints on sub-GeV dark matter candidates~\cite{Arnaud2018-nr}. A new spherical proportional counter with a gas volume of $O\left(m^{3}\right)$ is being developed, that will be operated under high pressure to further increase the target mass. The successful operation of the detector in high pressure and high gain relies heavily on the anode support structure, for response homogeneity, charge evacuation, and stability of operation.
The inclusion of a correction electrode, which in the design described above is just the grounded rod, is essential: It shields the volume of the detector from the anode wire and improves the electric field homogeneity for zenith angles near the wire~\cite{Giomataris2008-mx}. 

A second correction electrode could be placed a few millimetres from the anode surface, allowing the tuning of the electric field near the anode.
However, having two electrodes with an electric potential difference between them in close proximity increases the probability of a discharge, making the detector unstable. This can be a limitation for the operation of the detector at high gain under high gas pressures. The addition of layers of insulators has been proposed and tested but with undesirable side-effects due to charging-up.
In this article a different approach is presented, using electrodes made of a resistive material, specifically, lime-soda glass. Such materials have been used in other detectors to cover or act as electrodes~\cite{Naimuddin2014-mp,Kanishka2016-va} and helped to improve detector operation. 

\section{Electric field configuration}
\begin{figure}[htbp]
\centering
\subfigure[\label{fig:idealfield}]{\includegraphics[width=0.45\textwidth, angle =-0]{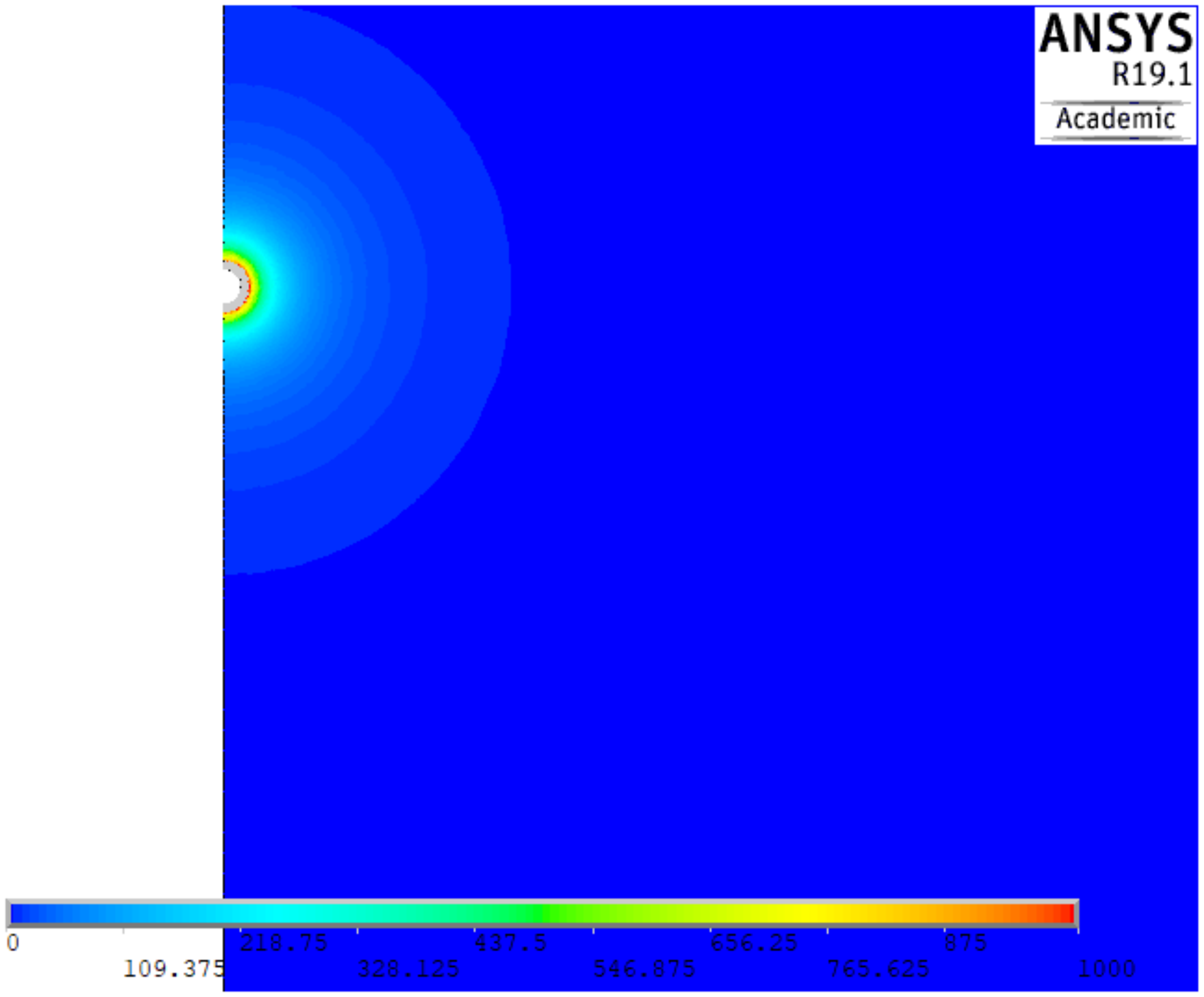}}
\subfigure[\label{fig:wirefield}]{\includegraphics[width=0.45\textwidth, angle =-0]{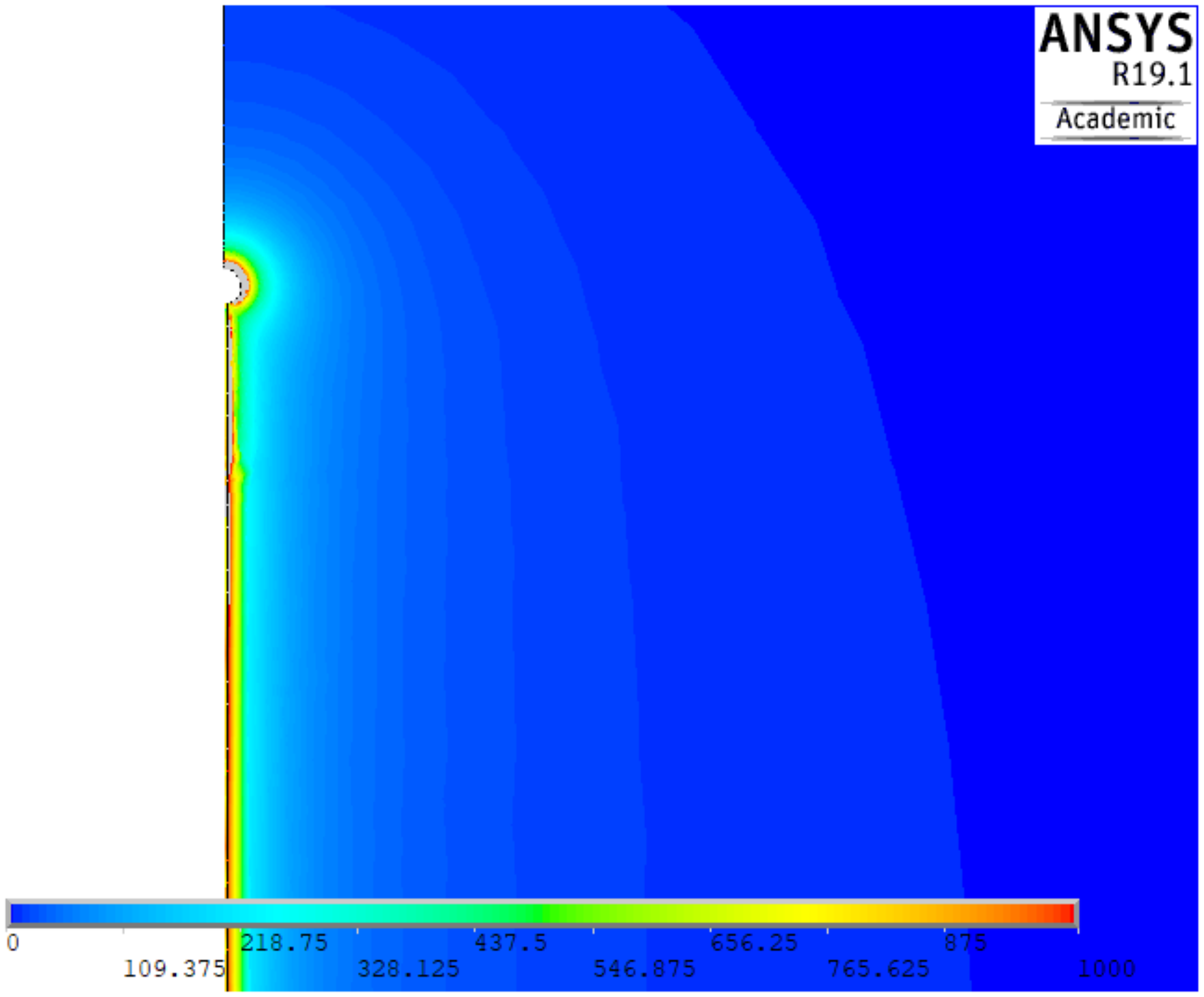}}
\subfigure[\label{fig:rodfield}]{\includegraphics[width=0.45\textwidth, angle =-0]{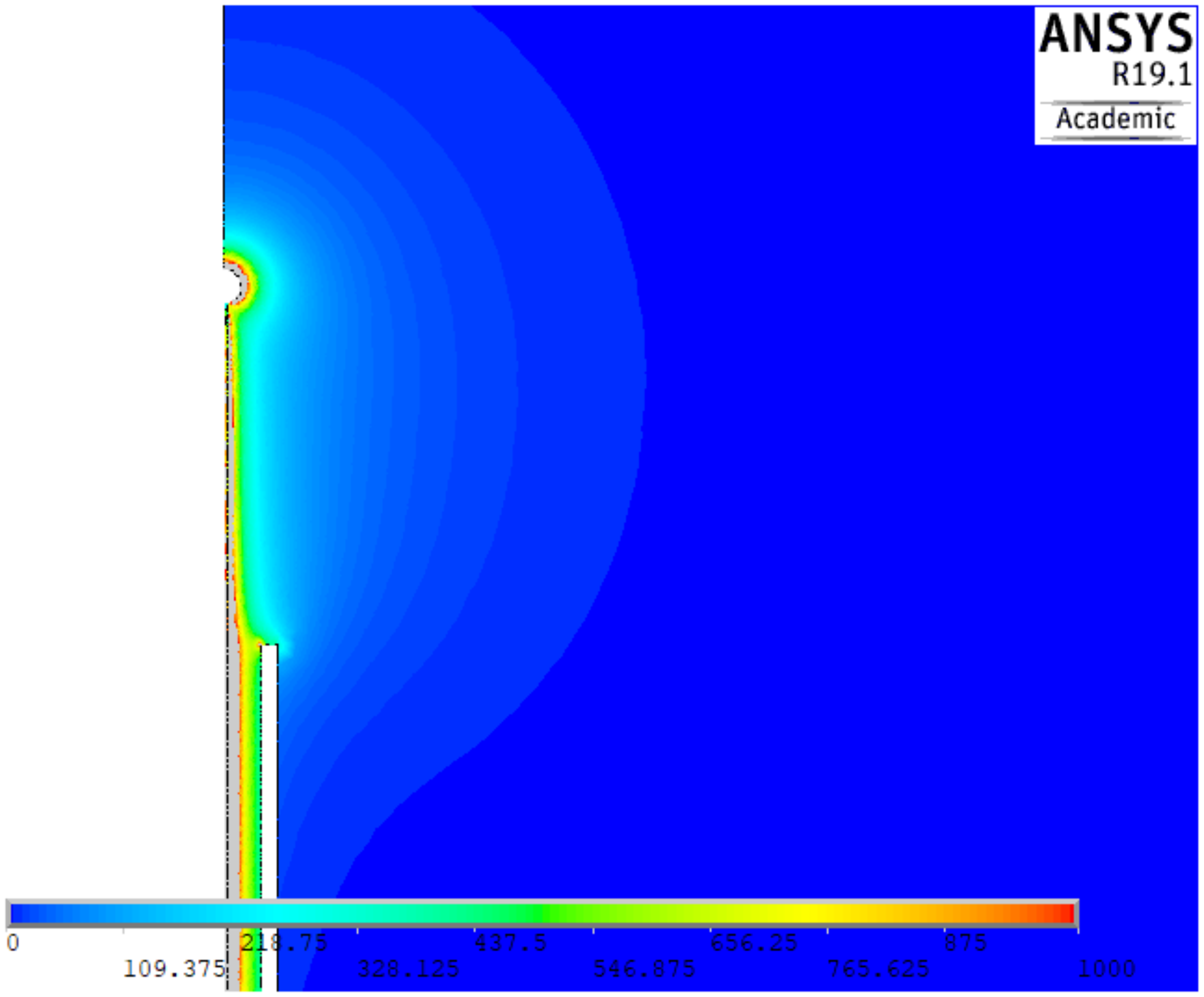}}
\subfigure[\label{fig:umbrellafield}]{\includegraphics[width=0.45\textwidth, angle =-0]{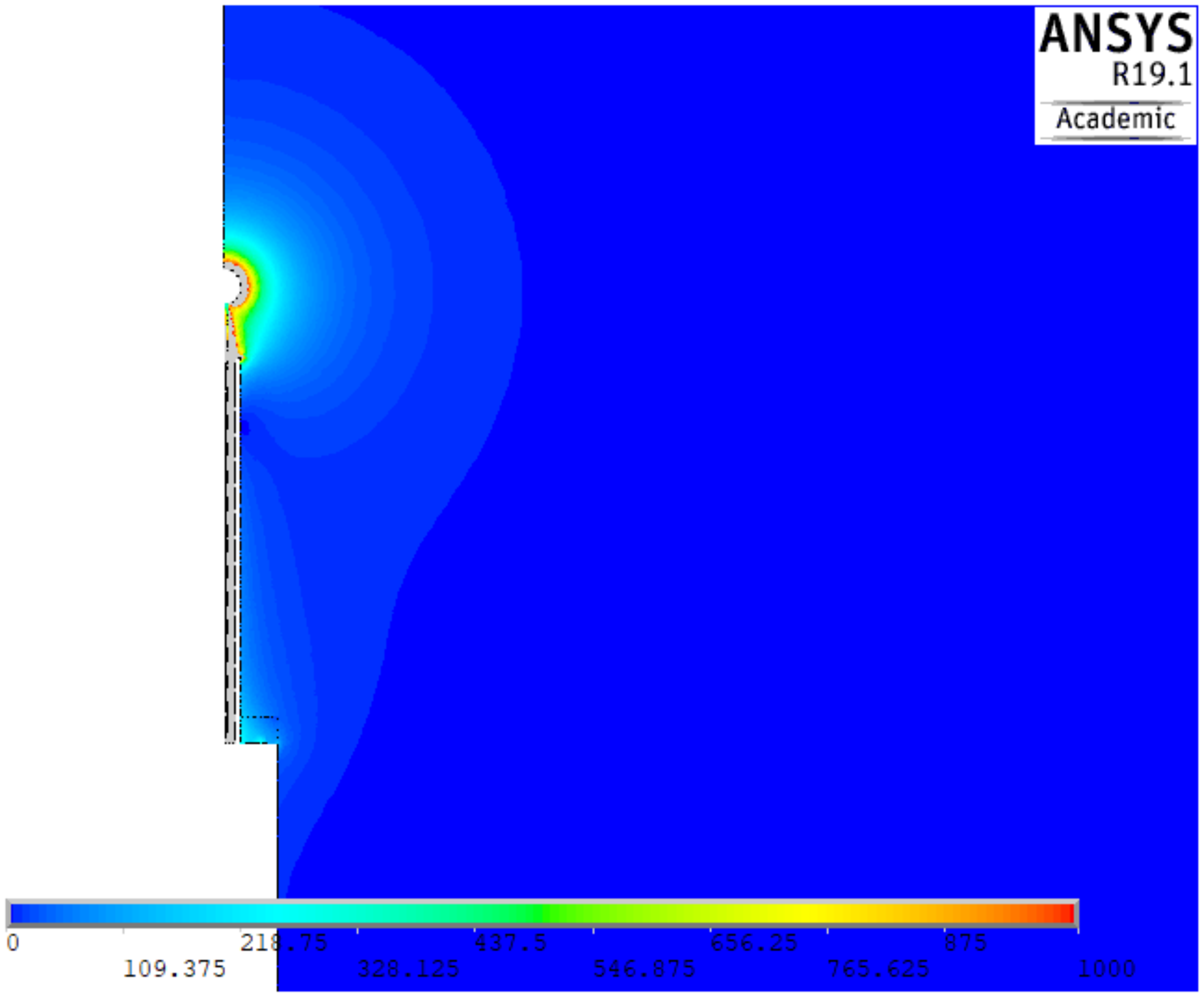}}
\caption{\label{fig:field}
Electric field equipotential lines calculated using FEM software for a $15~\si{\cm}$ radius spherical proportional counter with a $1~\si{\mm}$ radius anode at $2000~\si{\V}$  
\subref{fig:idealfield} for the ideal case comprising just the anode at high voltage,
\subref{fig:wirefield} including the anode wire to supply the high voltage, 
\subref{fig:rodfield} including the grounded metallic rod surrounding the anode wire, and \subref{fig:umbrellafield} including a second glass electrode placed $3~\si{\mm}$ from the anode and set to $250~\si{\V}$.
}
\end{figure}

The correction electrode plays an essential role for the electric field tuning in realistic detector geometries, affecting the uniformity of response of the detector as a function of the initial ionisation position. In its simplest  form, the correction electrode is the grounded metallic rod, which shields the gas volume from the electric field of the anode wire.
In the absence of the rod, a large fraction of the electric field lines would terminate on the wire, rather than the anode, resulting in the loss of ionisation electrons. 
The beneficial effect of the metallic rod is demonstrated by comparing the electric field configurations in Fig.~\ref{fig:wirefield} and~\ref{fig:rodfield}, obtained with Finite Element Method (FEM) software \cite{Rao2018-gx}, to the ideal case shown in Fig.~\ref{fig:idealfield}.
An extended analysis on the use of the metallic ground rod can be found in Ref.~\cite{Giomataris2008-mx}. 

\begin{figure}[htbp]
\centering 
\subfigure[\label{fig:comparison}]{\includegraphics[width=.45\textwidth]{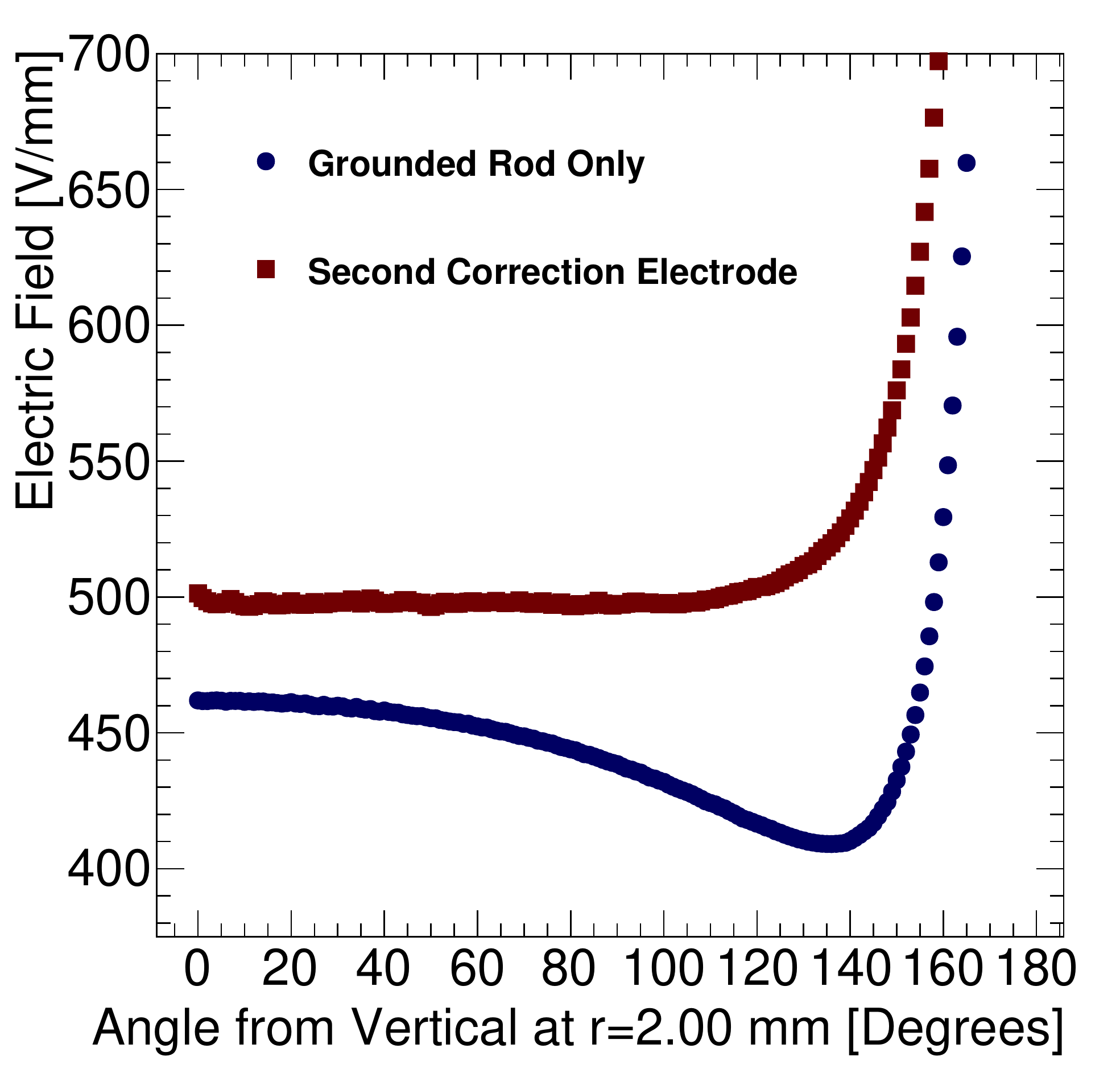}}
\subfigure[\label{fig:voltvar}]{\includegraphics[width=.45\textwidth]{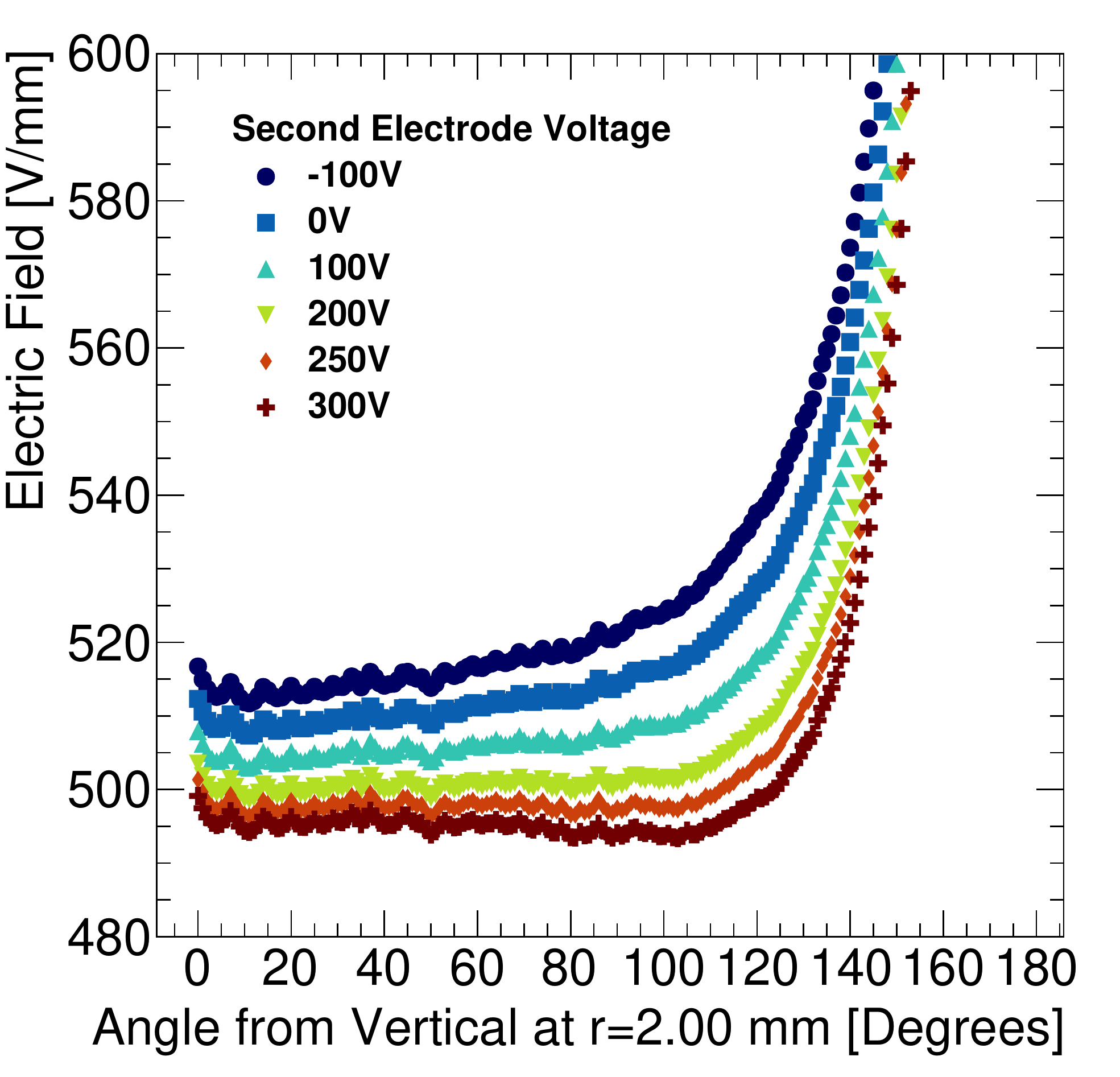}}
\caption{FEM calculation of the electric field as a function of the zenith angle calculated at a radius of $2~\si{mm}$ in an spherical proportional counter of radius $15~\si{cm}$ with a $1~\si{\mm}$ radius anode set at $2000~\si{\V}$. \subref{fig:comparison} The comparison of the electric field for the case of a grounded rod surrounding the wire to the anode and the case where a second electrode is included.
\subref{fig:voltvar} The electric field for a second electrode consisting of $20~\si{\mm}$ of glass located $3~\si{\mm}$ from the anode for various applied voltages on the second correction electrode. A voltage of approximately $250~\si{\V}$ provides the most homogeneous field for this geometry. 
}
\end{figure}

Nevertheless, a more homogeneous electric field through the volume of the detector and close to the surface of the anode, can be obtained by adding a second correction electrode, complementary to the grounded metallic rod. This electrode is thinner, $1~\si{\mm}$ to $2~\si{\mm}$ in diameter, and shorter than the rod and it is placed a few millimeters from the anode surface. An independent electric voltage can be applied to this electrode. The effect of the second correction electrode is shown in Fig.~\ref{fig:umbrellafield}, providing higher electric field homogeneity and, thus, uniform response for large zenith angles, as demonstrated in Fig.~\ref{fig:comparison}. The electric field around the anode can be fine-tuned by varying the voltage on the surface of the second correction electrode. This is particularly important for the hemisphere containing the grounded rod, as shown in Fig.~\ref{fig:voltvar}.

\section{The resistive correction electrode}
The correction electrode provides a simple but effective solution to the inhomogeneity of the electric field. However, to obtain the greatest control over the electric field it must be positioned only a few millimeters from the surface of the anode, supported by the grounded rod, which increases the probability of a discharge. 

Following initial conceptual designs using FEM software, several prototypes were constructed. Testing of these prototypes has demonstrated that it is impractical to implement such a correction electrode using metallic or any conductive materials due to the increased sparking rate and intensity. A solution explored was the coating of the conductive material in an insulating layer, usually consisting of a layer of glue, Araldite\textregistered, or PLO plastic. However, the introduction of insulating materials close to the anode induces charging-up of the electrode and results in time-dependent gain variations, with time constants depending on geometry and operational conditions. 

An alternative was found in the use of resistive materials as electrodes, which has been shown to reduce the spark rate and intensity in many detector designs, especially small gap micro-pattern detectors~\cite{Alexopoulos2011-xc,Wotschack2012-ly,Yoshikawa2012-eo}.
Furthermore, using resistive materials for electrodes is an effective technique to achieve higher gain operation and increased operational stability. The resistivity in these materials, is high enough to quench sparks and allow normal detector operation in the occasion of a spark, while it is adequately low to prevent charging-up and, thus, time-dependent gain variations.

A series of resistive materials has been tested. Following the experience of resistive bulk Micromegas, initial tests involved polymer paste with resistivity in the $10^{3} - 10^{6}\;\si{\ohm}\cdot\si{cm}$ range.
The construction of cylindrical electrodes at high precision when using paste was found to be challenging, while the obtained resistivity was inadequate for the desired effects. 

Glass as a resistive material has been used to construct  electrodes for various detectors including large surface Resistive Plate Chambers~\cite{Riegler2004-ts,Datar2009-co,Naimuddin2014-mp}. The glass electrode we used was in the form of a cylindrical tube with an $1.85(1.2)~{\rm mm}$ external (internal) diameter. The glass type was lime-soda glass~\cite{brittanica}, a common material used in a large number of applications.

\subsection{Resistivity of the glass electrode}
\begin{figure}[htbp]
\centering
\includegraphics[width=.45\textwidth]{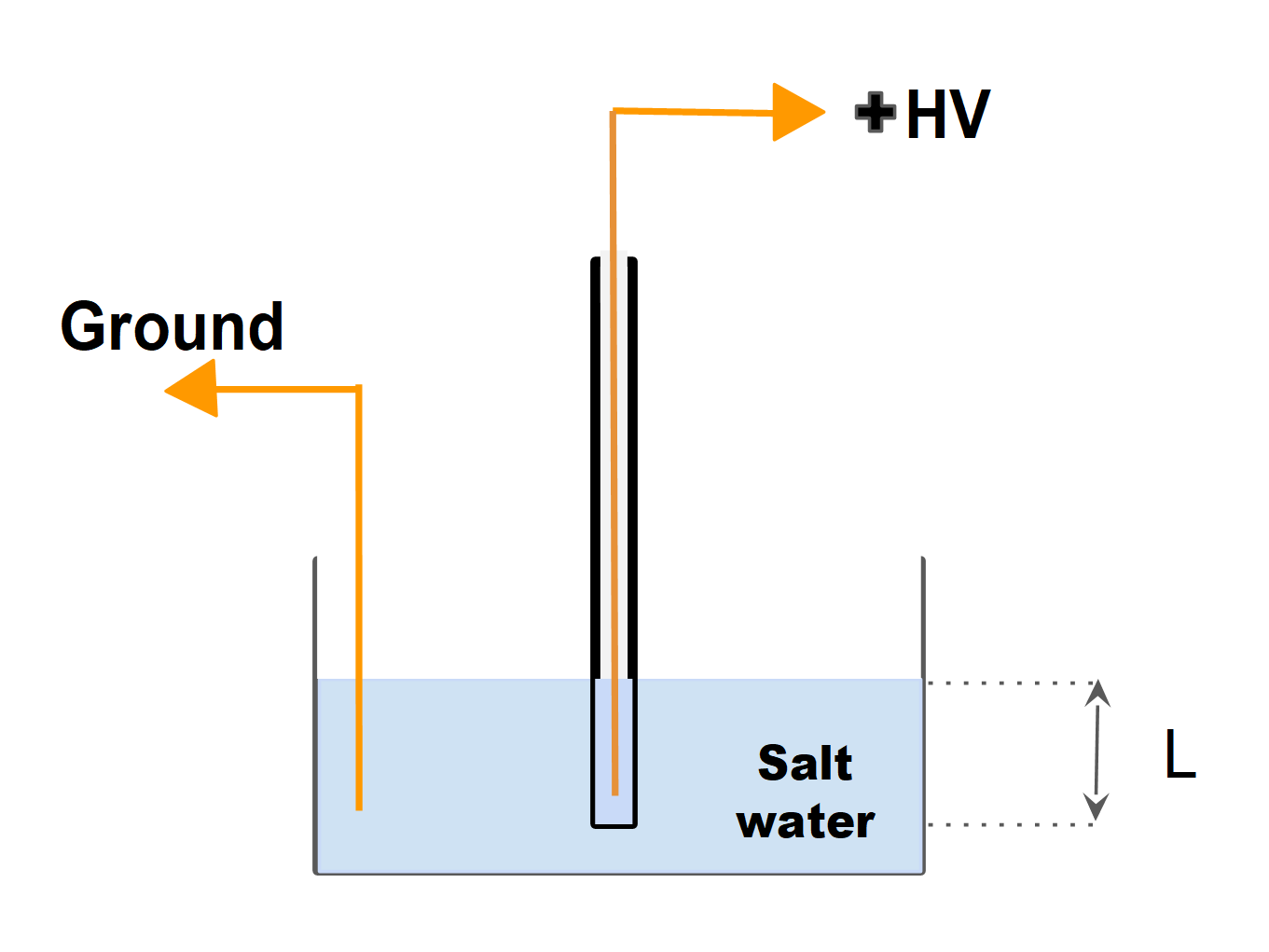}
\caption{Resistivity measurement setup: The glass tube is filled with and immersed in the solution to  depth $L$. A voltage is applied across the glass allowing measurement of the current through its volume.\label{fig:res_meas}
}
\end{figure}

\begin{figure}[htbp]
\centering 
\includegraphics[width=.45\textwidth]{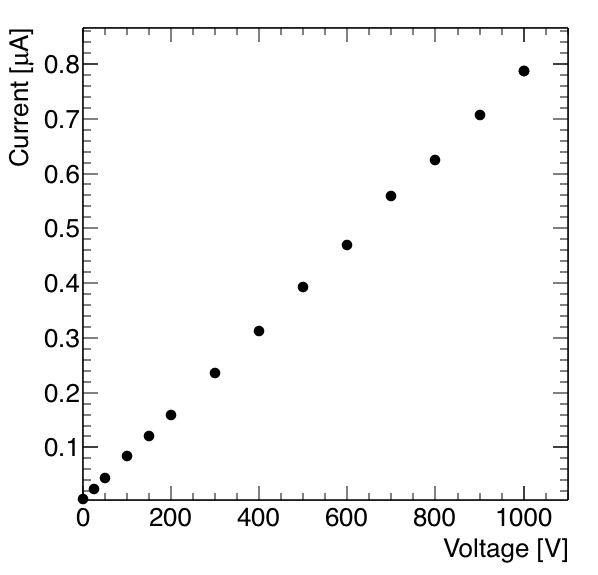}
\caption{
Current measured flowing through the glass versus applied voltage. From an average of these measurements, the resistivity of the glass tube was calculated as $\rho=5.05\cdot{10}^{10}\pm 27\%~\si{\ohm}\cdot\si{\cm}$
\label{fig:resistance}
}
\end{figure}

The volume resistivity, $\rho$, of the glass tube was measured based on a method described in Ref.~\cite{Coles1985-rf}. The principle of measurement is presented in Fig.~\ref{fig:res_meas}. The glass tube was sealed at one end with an insulating glue (Araldite) and heated at $200^{\circ}{\rm C}$ for a day to remove humidity from the surface. The tube was left to return to room temperature and was filled with an electrolyte of a salt water solution to a level $L$, shown in Fig.~\ref{fig:res_meas}. It was then immersed up to this level in a container filled with the same electrolyte. By using two copper wires, one immersed into the container and one in the tube, the electric circuit is closed and a voltage is applied.
Electrons must cross the volume of glass to be collected by the electrode inside the tube and measured. The current $I$ was measured as a function of the applied voltage $V$, up to $1~\si{\kV}$. A linear response is observed, as shown in Fig.~\ref{fig:resistance}.
The resistivity of the glass tube is calculated from the average of measurements using,
\begin{equation}
    \rho = \frac{2\pi LR}{ln(b/a)}\, ,
\end{equation}
where $R=V/I$ is the resistance between the outer and inner surface of the tube, $L=2.2~{\rm cm}$, $a=0.12~{\rm cm}$ ($b=0.185~{\rm cm}$) is the internal (external) diameter of the tube. The resistivity obtained is 
$\rho=5.05\cdot{10}^{10}\pm 27\%~{\rm \Omega}\cdot{\rm cm}$
. The uncertainty on the measured resistivity arises mainly from the limited accuracy of the electric current measurement in the voltage range between $0~\si{\V}$ and $150~\si{\V}$.

\begin{figure}[htbp]
\centering
\subfigure[\label{fig:sensor_des}]{\includegraphics[width=0.55\textwidth]{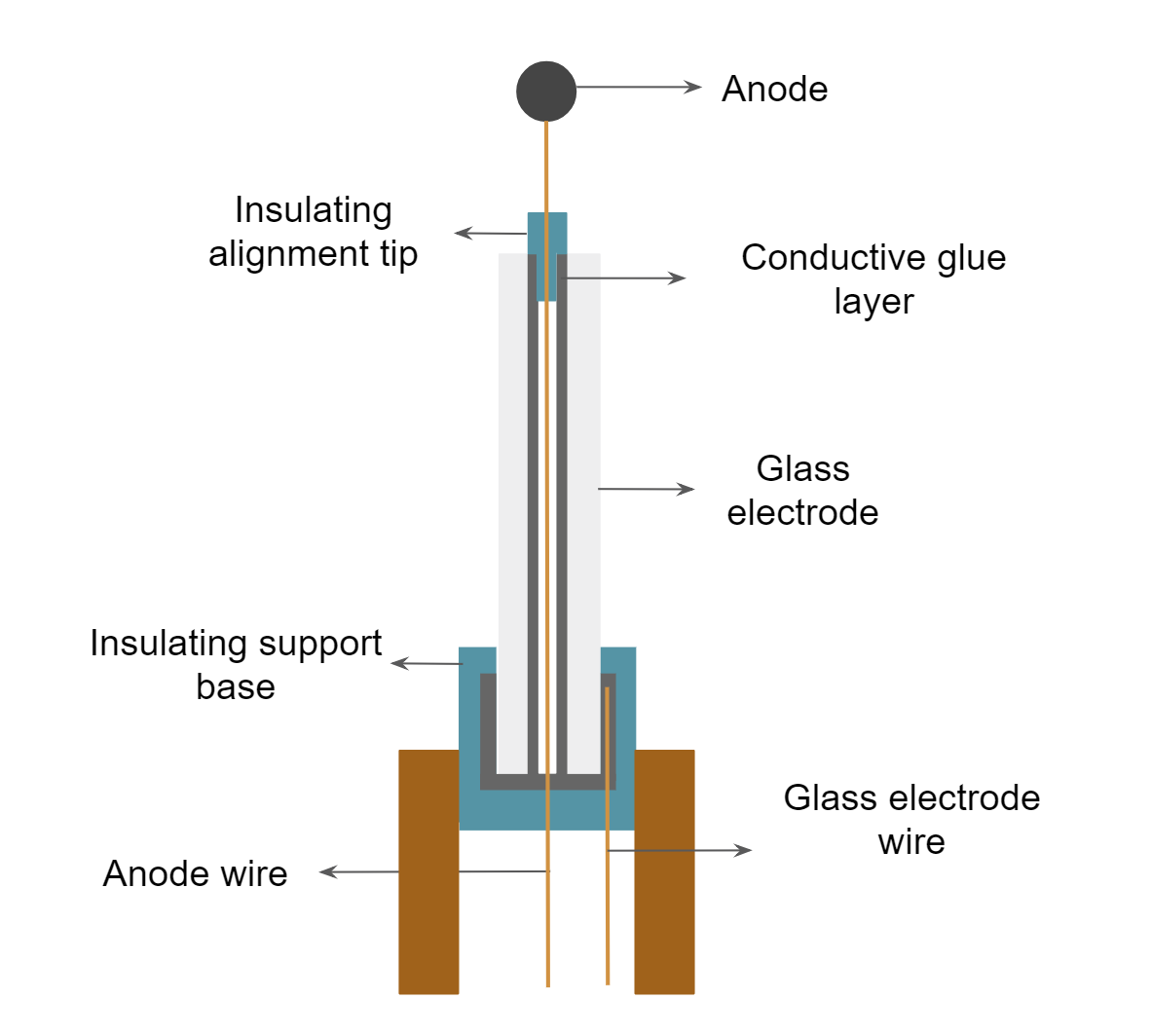}}
\subfigure[\label{fig:sensor_real}]{\includegraphics[width=0.4\textwidth]{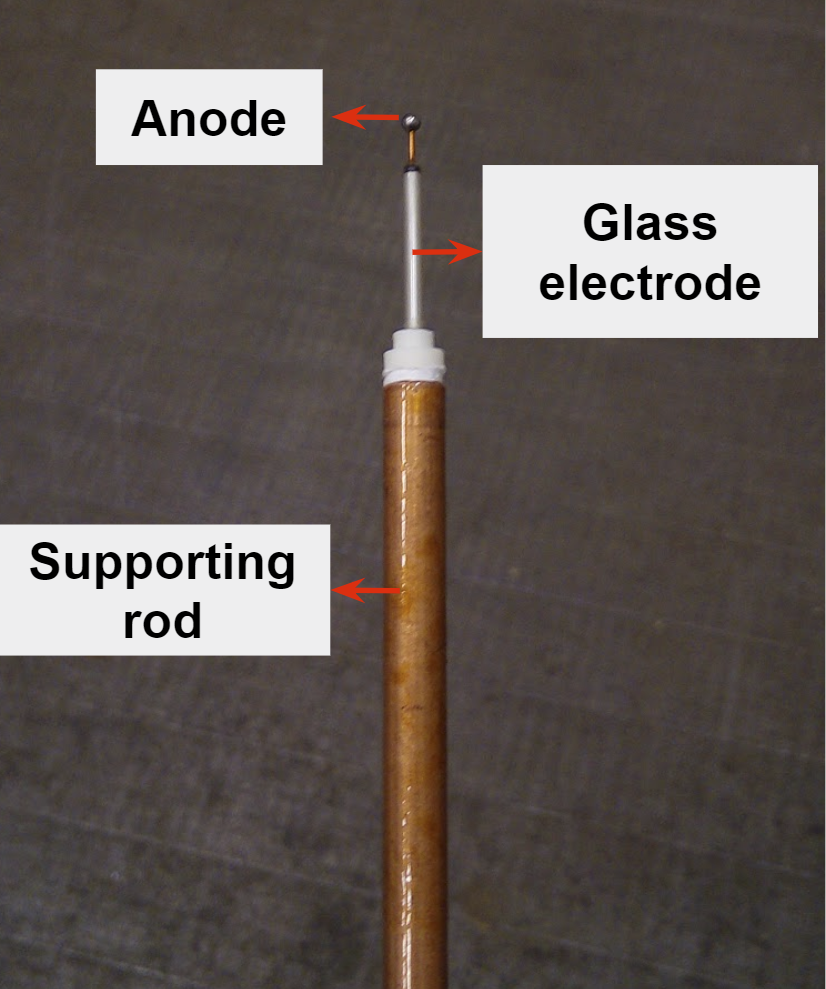}}
\caption{Module with a cylindrical glass correction electrode: \subref{fig:sensor_des} schematic and \subref{fig:sensor_real} implementation.
\label{fig:sensor} }
\end{figure}

\section{Development and performance of the resistive glass electrode prototypes}

The rod with the glass electrode structure is presented in Fig.~\ref{fig:sensor_des}. The internal surface of the glass tube is covered with a layer of conductive glue and electric potential can be applied through a thin wire. This conductive layer facilitates the application of the electric potential on the glass electrode and shields from the influence of the wire's electric field. The external part of the layer is covered with insulating glue, with a resistivity greater than ${10}^{14}$ $\Omega\cdot{\rm cm}$,
to avoid sparks between the conductive layer and the rod. The wire of the anode passes through the glass tube and is aligned by a short insulating tip. The bottom part of the glass tube with the conductive layer and the insulating cover is placed in a insulating support base, made of PLO, constructed with high precision using 3D-printing. Finally, the support base is inserted into the rod. 

In Fig.~\ref{fig:sensor_real}, the constructed module is presented.

The module is composed of a $2\;{\rm mm}$ diameter anode made of stainless steel. The anode is connected to the high voltage power supply with a thin wire, $150~\si{\um}$ in diameter, surrounded by a $200~\si{\um}$ thick insulating material.
The glass tube has a length of $20\;{\rm mm}$ and the distance between the ending edge of the tube and the surface of the anode is $3\;{\rm mm}$. The module is at the centre of a spherical, stainless steel vessel of $15~\si{\cm}$ radius and is supported by a copper rod with a $4~\si{\mm}$ ($6~\si{\mm}$) inner (outer) diameter. The connection interface between the rod and the detector's spherical vessel ensures that both be grounded. The spherical vessel is $3\;{\rm mm}$ thick and is built to sustain pressure up to $10\;{\rm bar}$.
The module was tested for sparking in an argon atmosphere. 
The secondary electrode was grounded and the anode voltage gradually increased to $7000~\si{V}$, without any sparks being observed.

In the following, tests of the performance of the $15~\si{cm}$ radius spherical proportional counter are presented. The counter was prepared by initially evacuating the spherical shell using a turbomolecular pump, to a pressure of $5.37\cdot{10}^{-5}~\si{mbar\cdot L}$.

\subsection{Voltage application on the glass electrode}

\begin{figure}[htbp]
\centering 
\includegraphics[width=.6\textwidth]{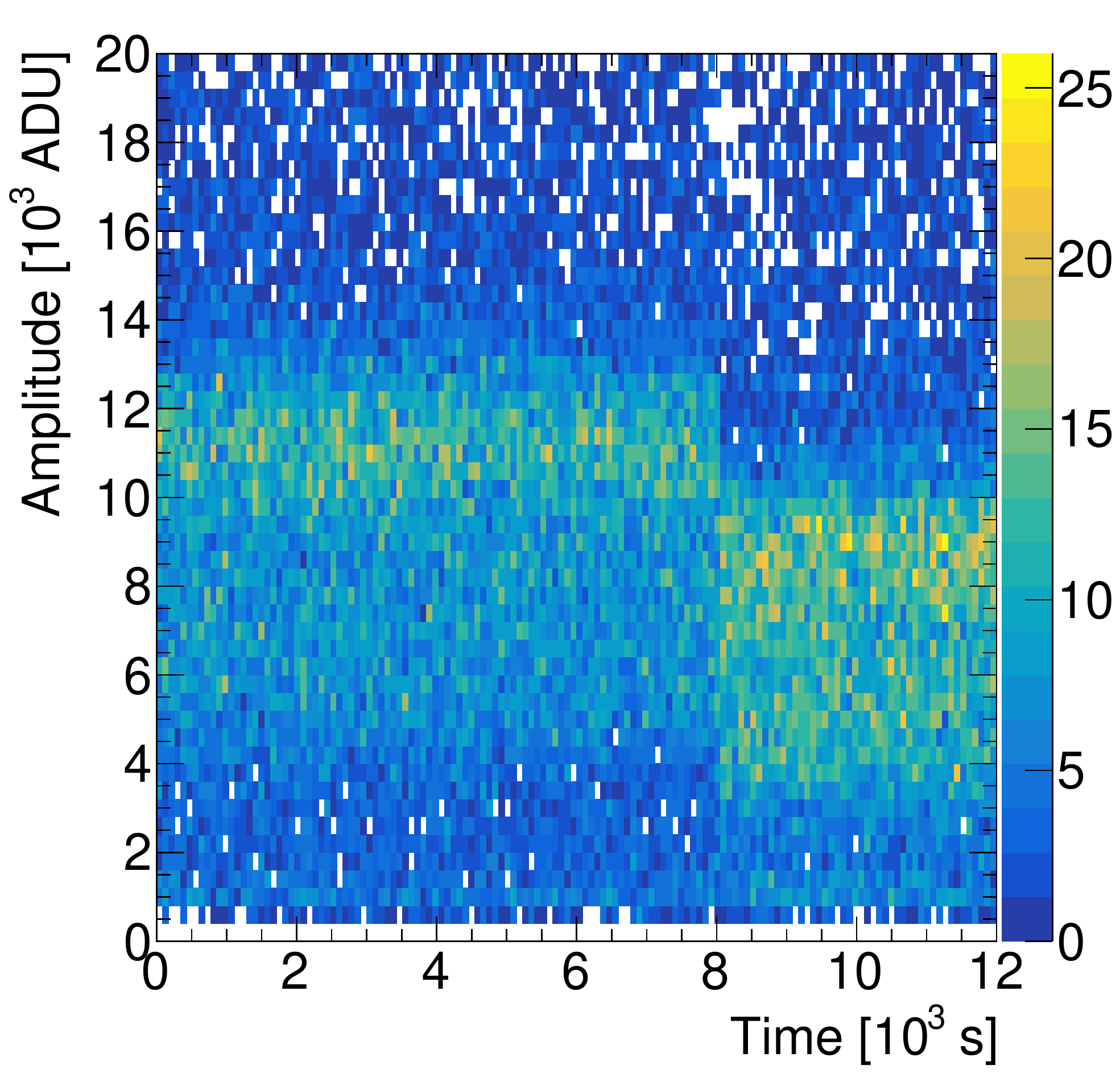}
\caption{Amplitude versus time in a module with a second correction electrode. 
Initially, the second correction electrode voltage was at $100~\si{V}$ but was increased to $200~\si{V}$ at $8000~\si{s}$. 
\label{fig:se07e000timeCorvsamp}
}
\end{figure}

A test was carried out to study the response of the module to a change in the voltage of the second correction electrode. The detector was filled with a gas of He:Ar:CH$_{4}$ (87\%:10\%:3\%) at a pressure of $1~\si{bar}$, with the anode voltage set to $1640~\si{V}$. The high voltage applied on the glass electrode was varied to check that it is properly applied on the electrode surface. The voltage variations resulted in simultaneous gas gain variations.
An example is displayed in Fig.~\ref{fig:se07e000timeCorvsamp} which shows the measured pulse amplitude as a function of time. Initially, the second correction electrode voltage was $100\l\si{V}$, but was changed to $200\;\si{V}$ after $8000\;\si{s}$, where the detector responded immediately with a gain drop. 

\subsection{Homogeneous response}
\begin{figure}[htbp]
\centering 
\includegraphics[width=.6\textwidth]{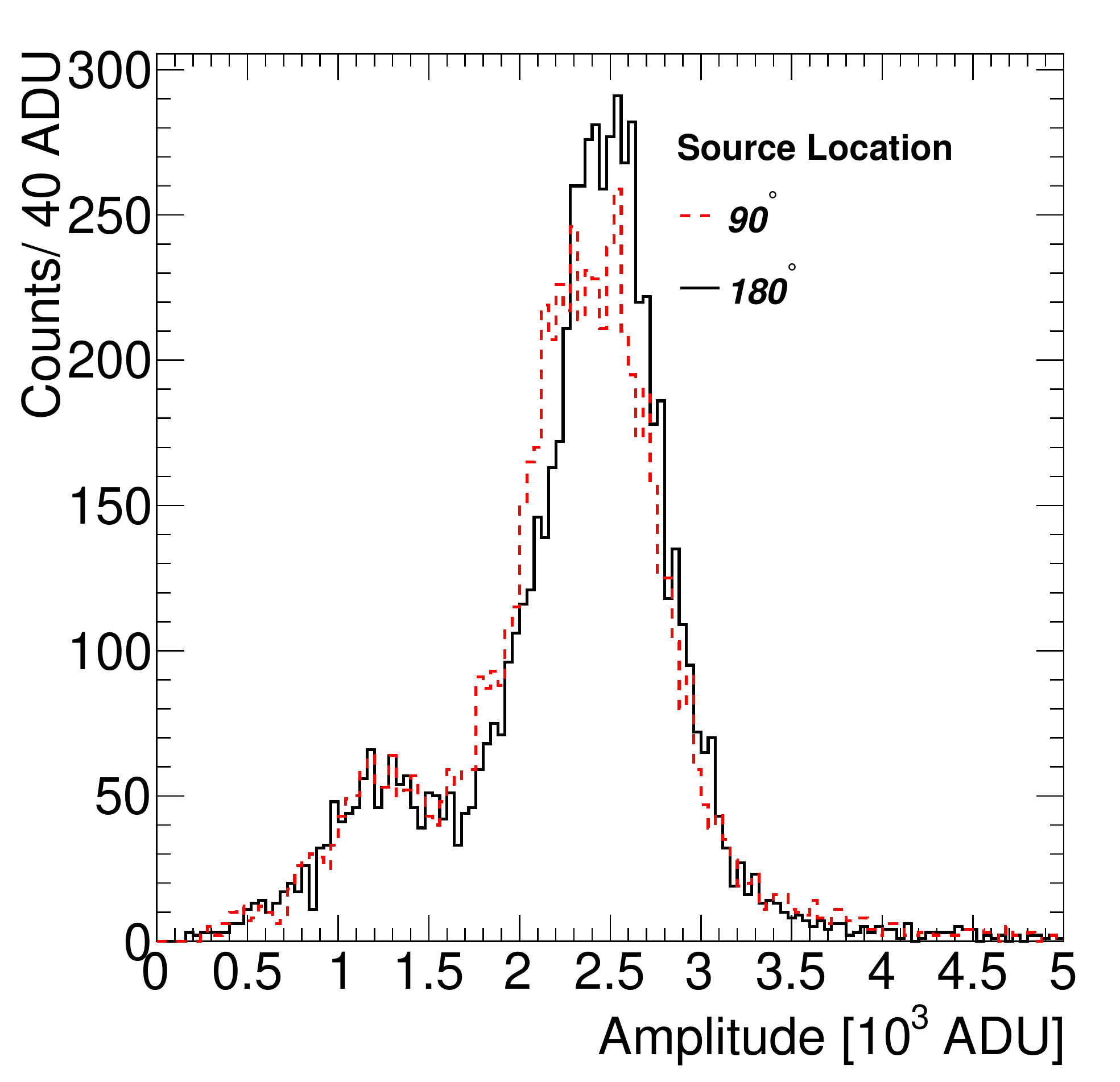}
\caption{The overlaid amplitude distributions for the recorded pulses, for $5.9~\si{keV}$ X~rays from an $^{55}$Fe source located inside the detector placed at a zenith angle of 
$\ang{90}$ (red) and $\ang{180}$ (black), relative to the grounded rod.
\label{fig:homogeneityFigures}
}
\end{figure}

To experimentally test the homogeneity of electric field achieved by the module, an $^{55}{\rm Fe}$ source was placed inside the detector. The location of this collimated source could be changed during detector operation. $^{55}{\rm Fe}$ decays through electron capture to $^{55}{\rm Mn}$ emitting $5.9~\si{keV}$ X~rays, on average. 
The detector was filled with $1~\si{bar}$ of He:Ar:CH$_{4}$ (92\%:5\%:3\%) with the anode and second correction electrode voltages set at $1450~\si{V}$ and $200~\si{V}$, respectively.
Data were collected with the source located at $\ang{90}$ and $\ang{180}$ to the grounded rod and their amplitude distribution is shown in Fig.~\ref{fig:homogeneityFigures}.
At $\ang{180}$ any distortions to the electric field caused by the wire and grounded rod should be minimised. However, at $\ang{90}$ the X~rays interact in a detector region potentially influenced by electric field distortions~\cite{Arnaud:2018bpc}. Such distortions could result in primary electrons arriving at the anode closer to the wire, leading to a worse energy resolution due to spatial gain variations. However, Fig.~\ref{fig:homogeneityFigures} demonstrate similar response in both cases. Thus, electric field distortions are corrected for by the second correction electrode.

\subsection{Operational stability}
\begin{figure}[htbp]
\centering 
\includegraphics[width=.6\textwidth]{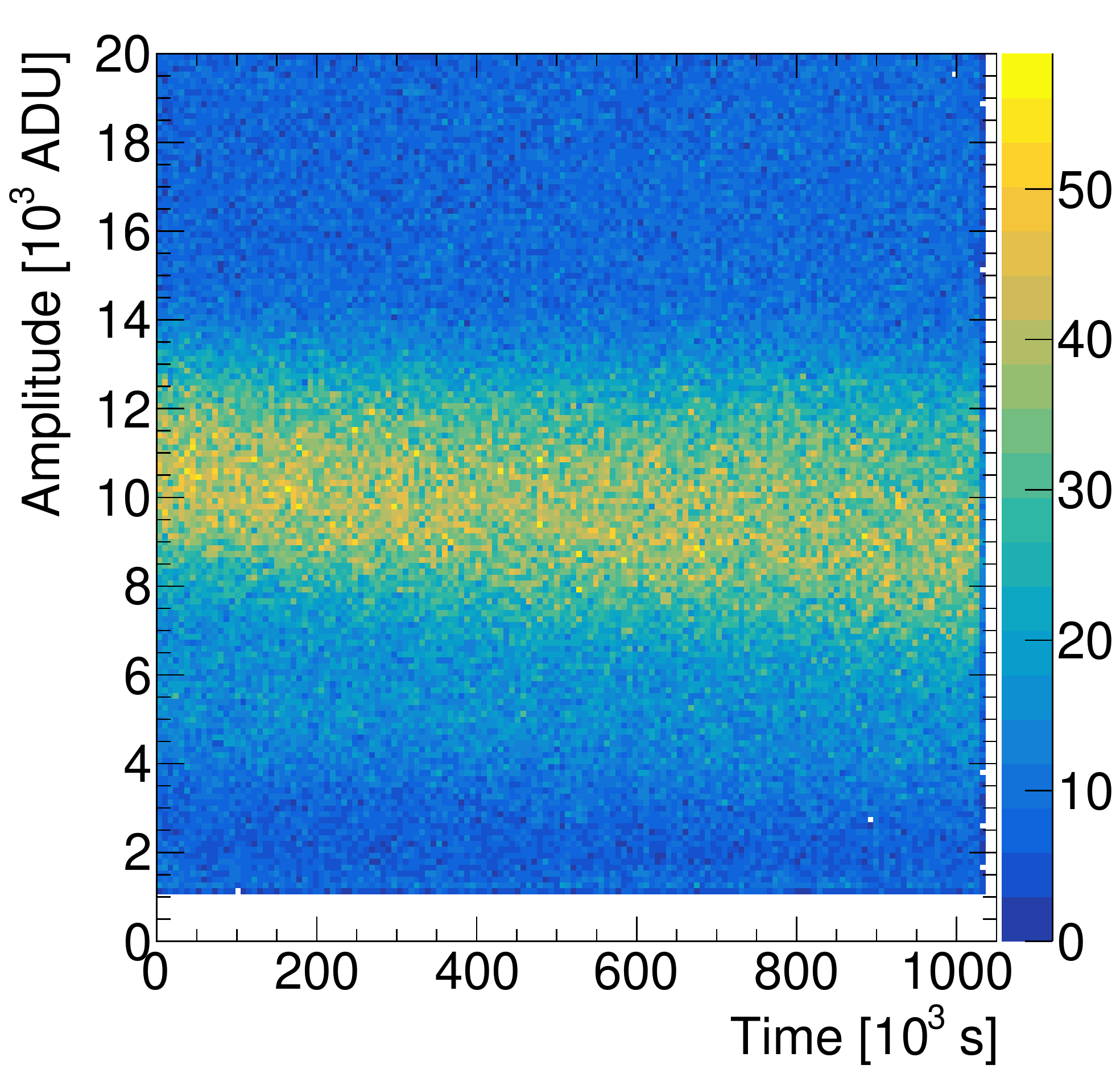}
\caption{
Pulse height as a function of time recorded using a module with a second correction electrode at $0~\si{V}$, with an anode voltage of $2350~\si{V}$ in a detector filled with $2~\si{bar}$ of He:Ar:CH$_{4}$ (87\%:10\%:3\%). The decrease in pulse height over time is due to contaminants, such as oxygen, leaking into the detector. Detector operation is stable and no discharges are observed.
\label{fig:stability}
}
\end{figure}

The main goal was to be able to operate the detector in a gas pressure up to 2 bar and to avoid sparks when voltages over  $2000~\si{V}$ were applied. To test the detector stability of operation, a gas of $2~\si{bar}$ of He:Ar:CH$_{4}$ (87\%:10\%:3\%) was introduced into the detector through an Oxysorb\textregistered\ filter to remove oxygen and water traces from the gas. A voltage of $2350~\si{V}$ was applied to the anode with $0~\si{V}$ applied on the second correction electrode. To monitor the change in gain over time, the $6.4~\si{keV}$ X~ray fluorescence of the $^{55}$Fe K-line was used \cite{Bougamont2010-xp}. This fluorescence X~ray is emitted homogeneously from the detector vessel stainless steel walls, induced mainly by environmental \textgamma-rays and cosmic radiation. Data were collected for $12$ days and the pulse height recorded versus time is shown in Fig.~\ref{fig:stability}. A gradual decrease of the pulse height is observed which is caused by the introduction of contaminants due to the imperfect vacuum tightness of the detector. However, the detector was stable throughout the entire period, with no spark-induced gain variations.

\section{Conclusions}
Thanks to its versatility and features, the spherical proportional counter has found its way to a series of applications involving detection and measurement of low energy recoils, X~rays, $\alpha$-particles, protons, $\gamma$-rays, and neutrons. Most of these applications require the detector to operate in high pressure with a relatively high gain, and long-term stability. Also good energy resolution and  homogeneous response is important for these applications. The use of a correction electrode, on which a second voltage can be applied, is an effective solution to achieve these goals. 
The proposed design featuring a glass resistive electrode provides a handle to fine-tune the electric field near the anode, optimise the resolution for given operation conditions, while suppressing sparks and ensuring long time operation stability. 

\acknowledgments

This work was performed within the NEWS-G collaboration and has received funding by the French National Research Agency (ANR-15-CE31-0008).


\bibliography{mybiblio}

\end{document}